\documentclass[aps,pra,letterpaper,twocolumn,amsmath,amssymb,floatfix]{revtex4-1}

\usepackage{graphicx}% Include figure files
\usepackage{dcolumn}% Align table columns on decimal point
\usepackage{bm}% bold math
\usepackage{epsfig}

\begin{document}

\title{Compact Toroidal Ion Trap Design and Optimization}

\author{M.J.~Madsen}
\author{C.H.~Gorman}
\affiliation{Department of Physics, Wabash College, Crawfordsville, IN  47933}

\date{\today}

\begin{abstract}
We present the design of a new type of compact toroidal, or ``halo'', ion trap.  Such traps may be useful for mass spectrometry, studying small Coulomb cluster rings, quantum information applications, or other quantum simulations where a ring topology is of interest.  We present results from a Monte Carlo optimization of the trap design parameters using finite-element analysis simulations that minimizes higher-order anharmonic terms in the trapping pseudopotential, while maintaining complete control over ion placement at the pseudopotential node in 3D using static bias fields.  These simulations are based on a practical electrode design using readily-available parts, yet can be easily scaled to any size trap with similar electrode spacings.  We also derive the conditions for a crystal phase transition for two ions in the compact halo trap, the first non-trivial phase transition for Coulomb crystals in this geometry.
\end{abstract}

\maketitle

\section{Introduction}

Recent interest in novel ion trap geometries stems from both the need for smaller, more compact radiofrequency (RF) Paul ion traps \cite{amini,ravi, crick, deslauriers,maiwald} and traps that have novel common modes of motion \cite{lin,kim,wunderlich}.  We present a new compact ``halo'' ion trap geometry based on toroidal ion traps \cite{schatz,lammert,austin} that might satisfy both of these needs.  This new geometry has the advantage of being much smaller (on the order of a few hundred microns in diameter) than previous toroidal traps.  Additionally, this geometry is very open both optically and electronically: ions trapped around the circular RF node interact via the Coulomb force across the circle with very little shielding, similar to ions in a Penning-type trap, giving rise to strong ion-ion interactions across the trap \cite{lupinski}.  To date, Coulomb crystals have been studied in a number of different RF trap geometries including a linear trap \cite{rafac,schiffer,dubin}, a spherical trap \cite{apolinario}, ions confined to a two-dimensional plane \cite{hasse}, ions in a large toroid trap with no ion interaction across the circle \cite{rahman,schweigert,schweigert2}, and ions in a Penning trap \cite{gilbert}.  Our compact halo trap geometry is complementary to many of these previously studied systems.  This geometry may also be suited to simulating small chemical rings \cite{wilson}, circular clusters of electrons on the surface of liquid helium \cite{leiderer}, and novel Ising models \cite{friedenauer}. 

We begin by defining by the key parameters and characteristics of the compact halo ion trap geometry.  Our setup is based on readily available electrode parts, making this configuration straightforward and inexpensive to implement.  We model both the RF and static potentials in the trap using finite element analysis; we present the results from a Monte Carlo optimization of the adjustable physical trap parameters yielding a trapping RF potential that is hyperbolic near the center of the trap.  %We might need to address the issue of the hyperbolic fields and ion trap community assumptions later.
In order to provide complete control of the ions, we also optimize offset static potentials that provide control over the trap aspect ratio.  In the final section we derive the conditions for a crystal phase transition for two ions in this new geometry.  We show that, unlike a linear trap, there is a non-trivial phase transition for only two ions.

\section{Halo Trap Geometry}

\begin{figure}[htbp]
\centering
\includegraphics[width=8cm,keepaspectratio]{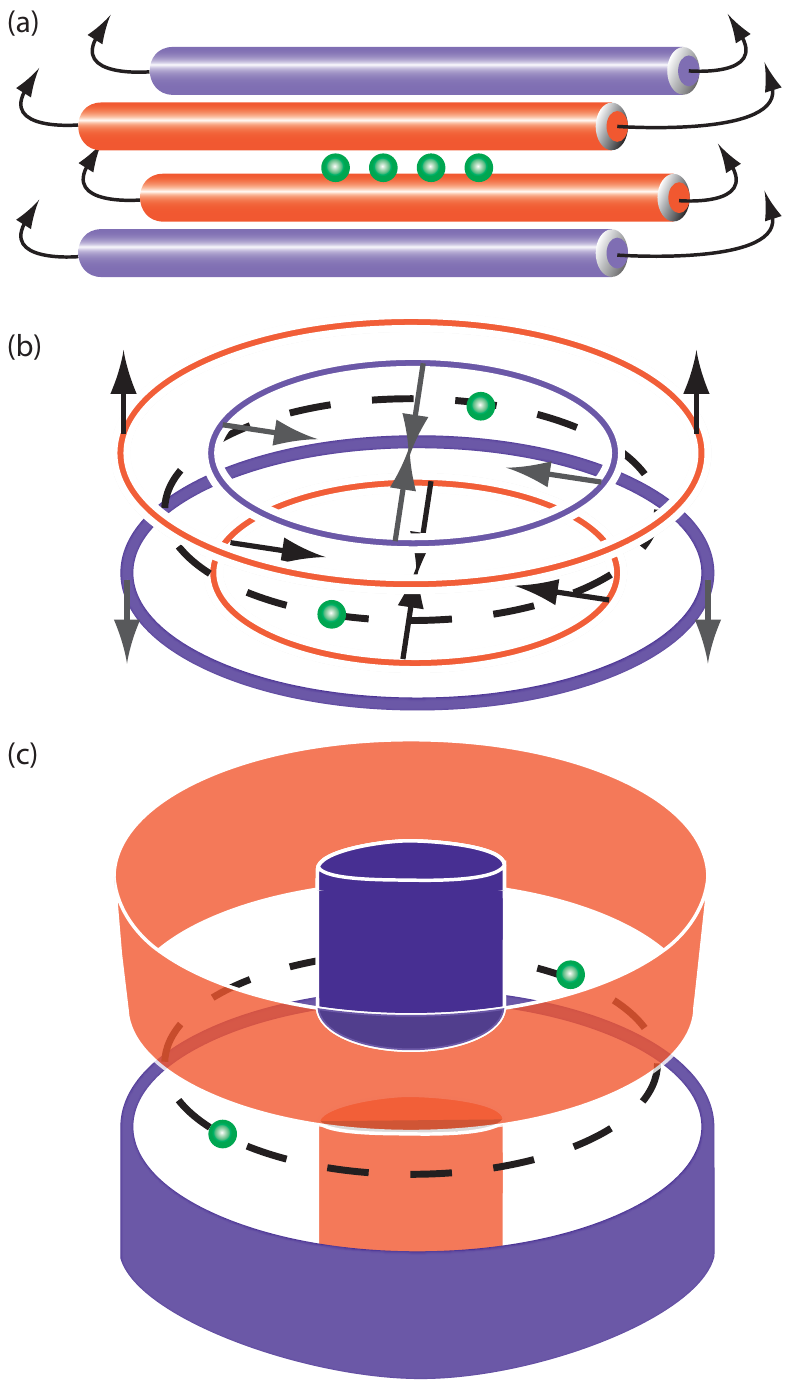}
\caption{(Color Online) (a) A standard four-rod RF linear ion trap is transformed into a toroid trap by wrapping the electrodes around onto themselves, forming the four rings of a toroid trap. (b) An RF toroid trap consists of four ring electrodes; RF potentials are applied to the top inner ring and the bottom outer ring.  This configuration is transformed into a compact halo ion trap by contracting the two inner rings down to cylindrical electrodes.  The outer electrodes are extended up and down for convenience in fabrication and assembly.  (c) The resulting form is our new halo trap design.  Radiofrequency potentials are applied to the top inner electrode and the bottom outer electrode, creating a circular RF node.}
\label{fig:HaloEvolution}
\end{figure}

A traditional RF quadrupole mass spectrometer or ion trap is created from four long cylindrical rods placed at the corners of a square as in Fig.~\ref{fig:HaloEvolution}(a).  Alternating RF potentials, $\pm V_0$, are applied to the electrodes forming a 2D quadrupole potential in the center of the square with a node that runs the length of the rods.  A toroidal RF trap is formed by wrapping the ends of linear trap around on themselves, forming four concentric loops stacked two on top of the others as in Fig.~\ref{fig:HaloEvolution}(b) \cite{schatz,austin}.  We propose one further modification that shrinks the toroidal trap and simplifies trap fabrication.  Since the inner rings are equipotential surfaces both on the top and the bottom, we shrink these rings down to solid, cylindrical electrodes, shown schematically in Fig.~\ref{fig:HaloEvolution}(c).  The outer two electrodes are also modified: we use conductive tubes to create these potentials, again concentric with the central electrodes.  We envision an ion trap with an RF node diameter of less than 1~mm.

\begin{figure}[htbp]
\centering
\includegraphics[width=8cm,keepaspectratio]{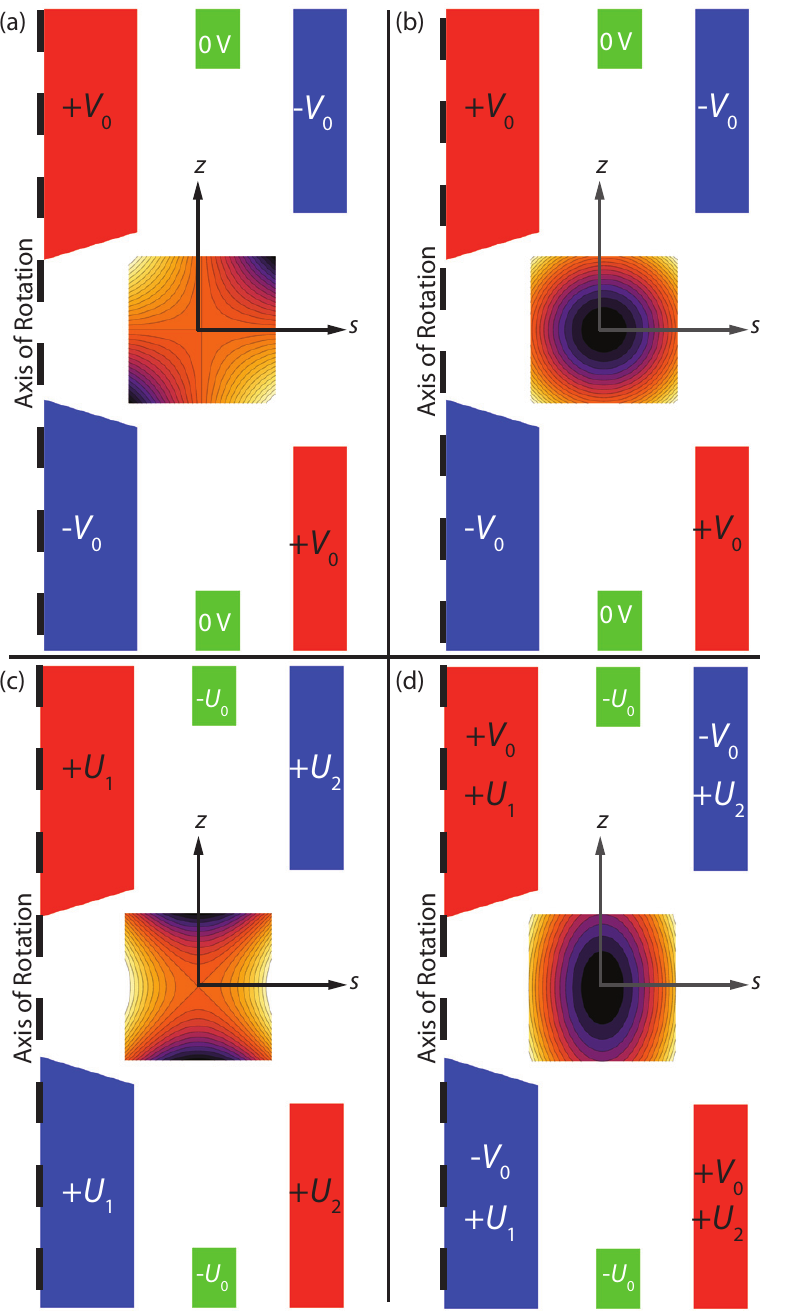}
\caption{ (Color Online) (a) The desired instantaneous RF potential is a quadrupole in the $s-z$ plane due to the voltages $\pm V_0$ applied to the RF electrodes.  Note that the middle control electrodes are held at RF ground. (b) The ideal pseudopotential is thus symmetric in the $s-z$ plane.  (c) In order to break the symmetry, static potentials $U$ are applied to the both the RF and control electrodes, forming a quadrupole rotated by 45 degrees from the RF potential.  (d) The combination of the static and pseudo potentials gives an asymmetric trap, indicating complete control over both the $s$ and $z$ trapping potentials.}
\label{fig:FieldSchematic}
\end{figure}

\begin{figure}[htbp]
\centering
\includegraphics[width=8cm,keepaspectratio]{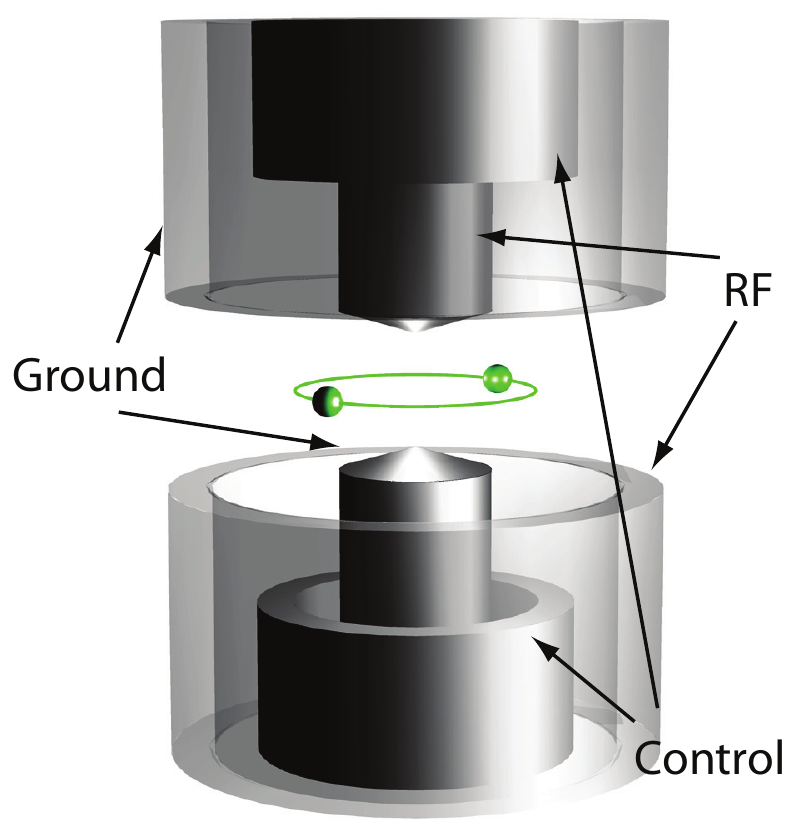}
\caption{(Color Online) Our halo trap design consists of three concentric cylindrical conductors.  The inner-most conductors are angled needle tips, the other two are cylindrical metal tubes.  The conductors are separated by insulators set back from the trap center (not shown). The inner and outer conductors form the RF trap; the middle conductor allows for static voltage control.}
\label{fig:Halo3D}
\end{figure}

In order to control the relative trap aspect ratio in the $s-z$ plane defined in Fig.~\ref{fig:FieldSchematic}, key to studying the crystal phase transitions \cite{rafac,schiffer,dubin}, we utilize both RF and static trapping fields.  Figure~\ref{fig:FieldSchematic} shows a cross-section of the halo geometry: rotating the $s-z$ plane about the axis located at the left of the figure yields the 3D trap configuration illustrated in Fig.~\ref{fig:Halo3D}.  The electrode geometry described above forms a quadrupole potential like that illustrated in Fig.~\ref{fig:FieldSchematic}(a) for potentials $\pm V_0$ applied to the center electrode and the outer tube.    The effective RF potential, or pseudopotential, is illustrated in Fig.~\ref{fig:FieldSchematic}(b), the center of which is located along the $z=0$ center line of the geometry.  To lowest order, the pseudopotential is symmetric in the $s-z$ plane; we discuss higher-order asymmetries below.  A static potential like that illustrated in  Fig.~\ref{fig:FieldSchematic}(c), which is rotated by 45 degrees from the RF potentials, is needed to control the relative trap strengths in the $s$ and $z$ directions.  We add a middle tube control electrode between the central cylinder and outer tube electrode that we can bias with static potentials to create this field.  This middle electrode is RF grounded with a static potential $-U_0$ applied to both the top and bottom electrodes.  Both the inner and outer electrodes are also biased with static potentials, $+U_1$ and $+U_2$, respectively.   The combination of the RF pseudopotential and this static potential will enable us to vary the trap aspect ratio, illustrated in Fig.~\ref{fig:FieldSchematic}(d).

There are a large number of free parameters available in the design and optimization of this halo trap geometry; we optimize a set of these parameters in order to minimize the higher-order, non-quadratic elements in the RF and static potentials.  We begin by fixing the inner and outer diameters of the cylinder and tubes that make the ion trap electrodes.  We choose stock parts (from Small Parts, Inc.) for these electrodes.  The inner cylinder is a stainless steel wire with an outer diameter of 510~$\mu$m (part GWX-0200).  The middle tube electrode is a 19 gauge stainless steel hypodermic round tube with an inner diameter (ID) of 830~$\mu$m and an outer diameter (OD) of 1100~$\mu$m (part HTX-19T).  The outer tube electrode is a 16 gauge stainless steel hypodermic round tube with an ID of 1350~$\mu$m and an OD of 1650~$\mu$m (part HTX-16T).  The electrodes are separated by concentric insulating tubes, matched to the various inner and outer diameters.  We located UHV-compatible stock parts made of polyimide that serve as the insulators (parts SWPT-028 and TWPT-045).  Although we choose specific parts for our design and optimization, all of our calculations serve for any size trap that has the same ratio of radii for the cylinder:(inner tube ID):(inner tube OD):(outer tube ID):(outer tube OD) ratio of 1:1.63:2.16:2.65:3.24.  

Having fixed the radii of the three electrodes, the number of adjustable free parameters available to optimize the potentials is now reduced to the spacing between the upper and lower sets of electrodes and the needle tip angle $\theta_n$ (shown in Fig.~\ref{fig:Parameters}).  We denote the spacings as $z_n$, $z_c$, and $z_t$ for the inner cylindrical ``needle'', the middle ``control'' electrode, and outer ``tube'' electrode, respectively, as described in Fig.~\ref{fig:Parameters}.   The tip angle will be fabricated by machining the inner wire prior to trap assembly.  The other dimensions described in the figure are the inner cylinder radius $R_n$, the radii of the two tubes $R_c$ and $R_t$ depicted as the distance from the center axis of rotation to the center of the electrode, and the tube wall thicknesses $t_c$ and $t_t$ for the control and outer tube electrodes.  These five dimensions are fixed by the electrode geometry described above.

We define three dimensionless parameters from these physical parameters and vary them to optimize the potentials.  First, the trap aspect ratio $A_h$ is defined as the ratio of the average $z$ separation of the needle and tube to the tube radius: $A_h \equiv (2 z_n + 2 z_t)/(2 R_t)$.  This roughly corresponds to the overall aspect ratio of the trap, or the distance between the top and bottom electrode structures.  The second parameter is the ``keystone'' $K_h$ defined as the ratio of the outer tube separation over the needle separation: $K_h \equiv 2 z_t/ 2 z_n$.  This parameter describes how much farther apart the outer tubes are compared to the inner cylinder.  The final parameter is defined as $V_h \equiv (z_c - z_t)/z_n$ and describes the control electrode separation as compared to the tube separation in units of the needle separation.  These parameters define several possible configurations simply: if all three electrodes are separated by the same amount equal to the tube radius $R_t$, the parameter set would be: $A_h = 2, K_h = 1, V_h = 0$.  A parameter set of $A_h = 1, K_h = 1, V_h = 1$ corresponds to separations of $z_n = R_t/2$, $z_c = R_t$, and $z_t = R_t/2$.  The geometry illustrated in  Fig.~\ref{fig:Halo3D} corresponds to a parameter set of approximately $A_h = 0.7$, $K_h = 1.7$, $V_h = 2$, and a needle angle of $17^\circ$. 

\section{RF and Static Potential Model}

\begin{figure}[htbp]
\centering
\includegraphics[width=8cm,keepaspectratio]{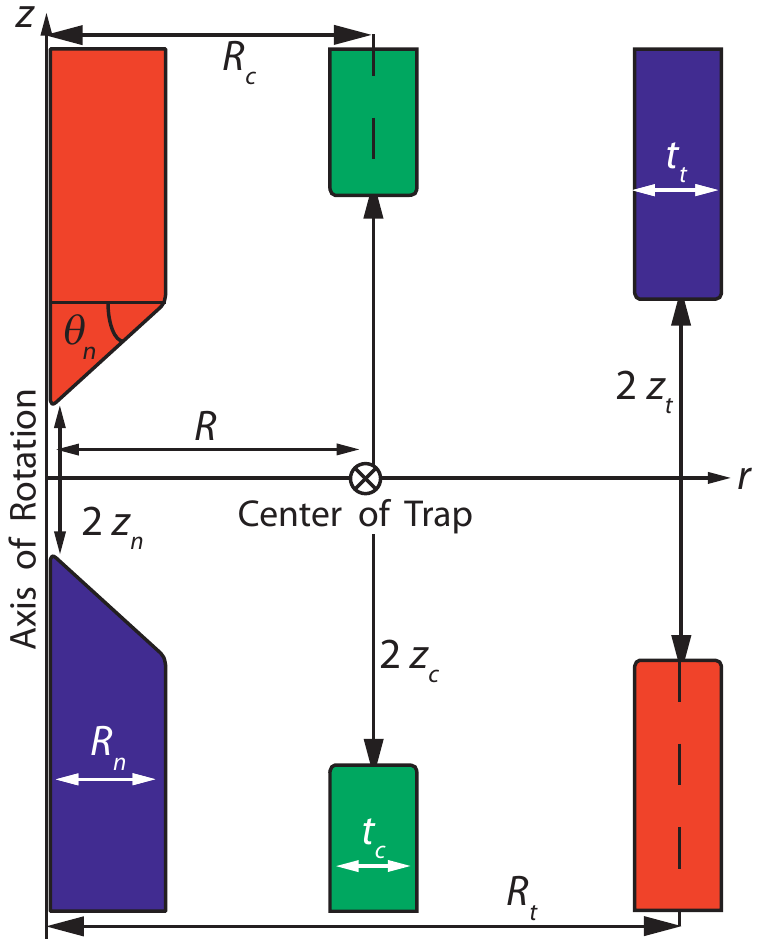}
\caption{(Color Online) There are four key dimensionless parameters that define the geometry of the halo trap electrodes: the needle angle $\theta_n$, the trap aspect ratio $A_h\equiv(2 z_n + 2 z_t)/(2 R_t)$, the trap keystone $K_h\equiv (2z_t)/(2z_n)$, and the control electrode offset $V_h\equiv (z_c-z_t)/z_n$.  The other dimensions are fixed and are properties of the conductors and insulators used in the model.}
\label{fig:Parameters}
\end{figure}

Although motion of ions in an RF Paul trap can be described in a variety of ways, the typical approach is to model the trapping potentials as an ideal quadrupole field \cite{wineland}.  The oscillating trapping potential $\Phi(r,z,t)$ for a toroidal trap in cylindrical coordinates $r$ and $z$ can be written as a spatially varying potential $V(r,z)$ which oscillates at the trap frequency $\Omega_T$: $\Phi(r,z,t) = V(r,z)\cos\Omega_Tt$.  The spatial component of the potential is typically described as a quadrupole field, $V(r,z) = V_0/r_N^2 \left(z^2 - r^2\right)$ where $V_0$ is the potential applied symmetrically to hyperbolic electrodes a distance $r_N$ from the trap center.  A toroidal trap like this halo trap shifts the trap center radially a distance $R$ from the $z$ axis as shown in Fig.~\ref{fig:Parameters}.  Although the purely quadratic field described by shifting the quadrupole field $V(r,z) = V_0/r_N^2 \left(z^2 - (r-R)^2\right)$ is not possible, we are interested a trapping potential where the quadratic term is dominant \cite{lammert}.  Thus, we fit the actual halo trap potential to the model hyperbolic potential $V(r,z) = V_0/\ell_\textrm{RF}^2 \left(z^2 - s^2\right)$, where $s = r-R$, and $\ell_\textrm{RF}$ is a single fit parameter.  However, since the RF fields are rotated by 45 degrees from the $s-z$ axes, as seen in Fig.~\ref{fig:FieldSchematic}(a), we rotate the coordinate system of the hyperbolic potential in order to compare it directly with the halo trap field.  The quadratic potential model describing the field in Fig.~\ref{fig:FieldSchematic}(a) is thus
\begin{equation}
V_\textrm{RF}(s,z) = -\frac{2 V_0}{\ell_\textrm{RF}^2} s z.
\label{eqn:RFideal}
\end{equation}
The static control potential shown in Fig.~\ref{fig:FieldSchematic}(c) can also be modeled as a hyperbolic potential.  Although we apply different control potentials to the needle and tubes, we fit the actual static potential with a single effective (unit) potential $U_\textrm{eff}$ and the fit parameter $\ell_\textrm{static}$.  The corresponding static potential model is then
\begin{equation}
U_\textrm{static}(s,z) = \frac{U_\textrm{eff}}{\ell_\textrm{static}^2}\left(s^2 - z^2\right).
\label{eqn:DCideal}
\end{equation}
We calculate the actual RF and static potentials, $V_\textrm{CH}(r,z)$ and $U_\textrm{CH}(r,z)$, for this compact halo (CH) geometry for a given parameter set using finite element analysis.  We use single-parameter fits (with $V_0=1$~V and $U_\textrm{eff}=1$~V) to find the normalization distances $\ell_\textrm{RF}$ and $\ell_\textrm{static}$ for the RF and static fields.  We then define a $\chi^2$ metric for the two-dimensional scalar fields to determine the goodness of the fit.  We integrate the square of the difference between the real and hyperbolic potentials over a circle area $A$ covering the center of the trapping region,
\begin{align}
\chi_\textrm{RF}^2 & = \iint\limits_A \left(V_\textrm{RF}(s,z) - V_\textrm{CH}(s,z)\right)^2 ds dz \\
\chi_\textrm{static}^2 & = \iint\limits_A \left(U_\textrm{static}(s,z) - U_\textrm{CH}(s,z)\right)^2 ds dz.
\end{align}
We minimized both of these metrics simultaneously through an iterative Monte Carlo approach described below.

\begin{figure}[tbp]
\centering
\includegraphics[width=8cm,keepaspectratio]{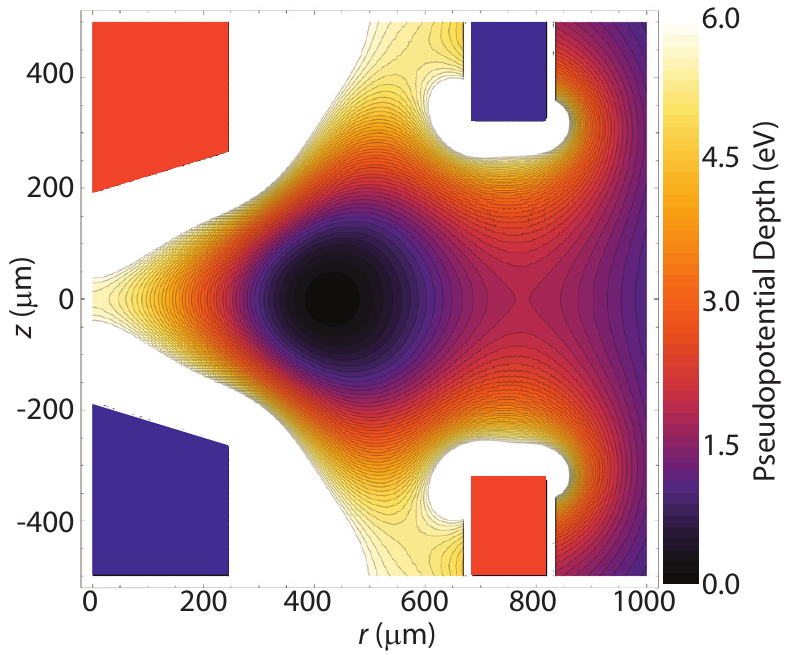} 
\caption{(Color Online) The pseudopotential $\psi$ for the optimized compact halo trap has a trap depth constrained by the saddle point along the $z=0$ symmetry line.  The pseudopotential is shown in electron-volts for a $^{24}$Mg$^+$ ion with an applied potential of $V_0=300$~V, and a trap frequency of $\Omega_T = 2 \pi \times80$~MHz.}
\label{fig:depth}
\end{figure}

We also investigate the depth of the RF trap, a particular concern for small and novel trap geometries \cite{deslauriers,maiwald}.  The depth is found by calculating the pseudopotential $\psi(r,z)$ from the calculated RF trap potentials \cite{dehmelt}:
\begin{equation}
\psi(r,z) = \frac{q^2}{4 m \Omega_T^2}\left|\nabla V_\textrm{CH}(r,z)\right|^2
\end{equation}
where $q$ and $m$ are the ion charge and mass.  There is a saddle point in the compact halo pseudopotential located along the $r$ axis approximately between the outer control electrode.  The trap depth in eV is shown in Fig.~\ref{fig:depth} for an applied voltage of $V_0=300$~V, $^{24}$Mg$^+$ ions, and a trap frequency of $\Omega_T = 2 \pi \times80$~MHz.

\section{Halo Trap Optimization}

We optimize the RF and static potentials using the following Monte Carlo procedure:

\begin{enumerate}
\item Optimize the RF potential.
\begin{enumerate}
\item Choose initial values for the adjustable geometric parameters $A_h$, $K_h$, $V_h$, and $\theta_n$.
\item Calculate the RF potential using finite element analysis.
\item Fit the RF potential to the model and find $\ell_\textrm{RF}$.
\item Calculate the $\chi^2_\textrm{RF}$ goodness of fit.
\item Adjust all of the parameter values and then repeat steps (b) through (d) until $\chi^2_\textrm{RF}$ is minimized.
\end{enumerate}
\item Optimize the static potential.
\begin{enumerate}
\item Locate the trap center $R$ along the $r$-axis.
\item Calculate the static potential using finite element analysis using the optimized geometric parameters.
\item Adjust the static potentials $U_0$, $U_1$, and $U_2$ until the static potential saddle point lies at $R$ along the $z=0$ axis.
\item Fit the static potential to the model and find $\ell_\textrm{static}$.
\item Calculate the $\chi^2_\textrm{static}$ goodness of fit.
\end{enumerate}
\item Iterate the entire process until both $\chi^2_\textrm{RF}$ and  $\chi^2_\textrm{static}$ are minimized.
\end{enumerate}

\begin{figure}[htbp]
\centering
\includegraphics[width=8cm,keepaspectratio]{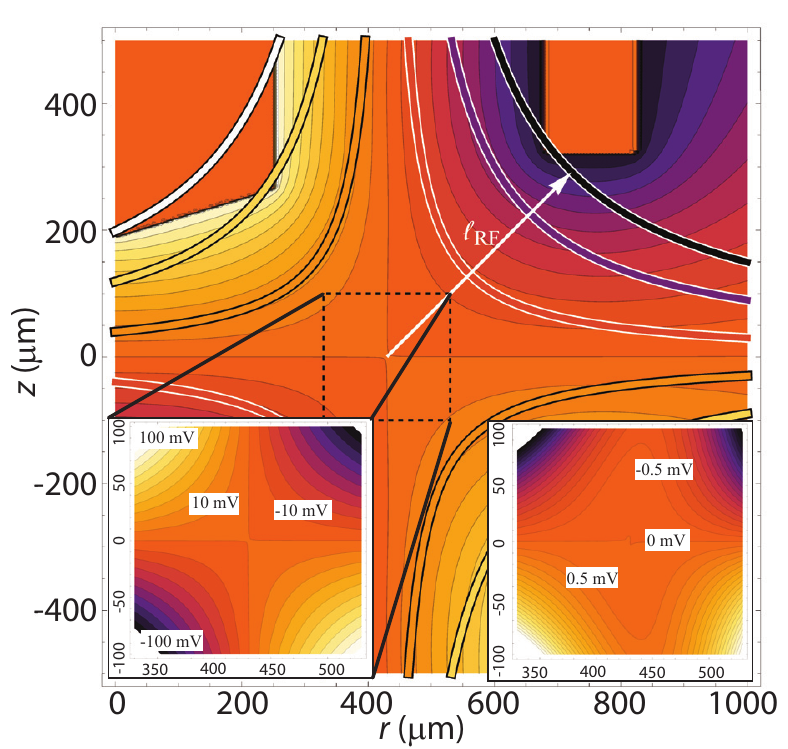}
\caption{(Color Online) The optimized RF instantaneous quadrupole potential $V_\textrm{CH}$ (with $\pm1$~V applied to the needle and outer ring electrodes) across the trapping region.  The ideal quadrupole fit to the potential has a normalization distance of $\ell_\textrm{RF}\approx413$~$\mu$m.The bottom left inset shows the potential near the center of the trap with contour lines separated by 10~mV.  The bottom right inset shows the difference between $V_\textrm{opt}$ and the ideal quadrupole potential $V_\textrm{ideal}$ with contours spaced at 0.5~mV.  The central flat region is approximately 100~$\mu$m long by 100~$\mu$m wide, much larger than typical ion excursions from the RF node in an ion trap.}
\label{fig:RF_pot}
\end{figure}

\begin{figure}[htbp]
\centering
\includegraphics[width=8cm,keepaspectratio]{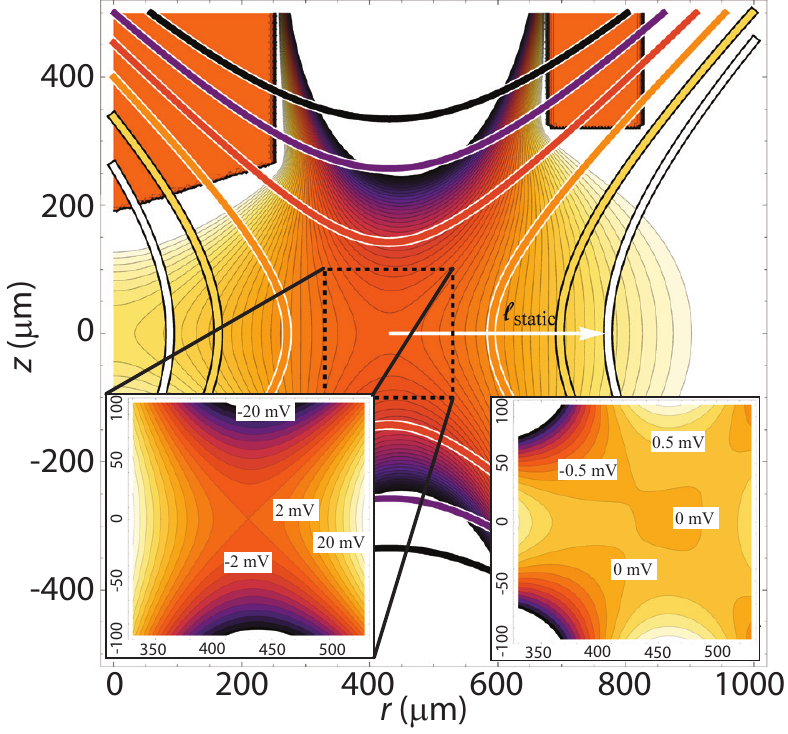}
\caption{(Color Online) The optimized static quadrupole potential $U_\textrm{CH}$ across the trapping region.  The ideal quadrupole fit to the potential has a normalization distance of $\ell_\textrm{static}\approx328$~$\mu$m.  The bottom left inset shows the potential near the center of the trap with contour lines  separated by 2~mV.  The bottom right inset shows the difference between $U_\textrm{CH}$ and the ideal quadrupole potential $U_\textrm{static}$ with contours spaced at 0.5~mV.  The central flat region is approximately 100~$\mu$m long by 100~$\mu$m wide, much larger than typical ion excursions from the RF node in an ion trap.}
\label{fig:DC_pot}
\end{figure}

Since the saddle points for the RF and static potentials do not necessarily have the same spatial location, unlike traditional four-rod traps, we simultaneously optimize both the RF and static potentials under the constraint that the saddle points overlap along the $r$-axis.  The optimized RF and static fields are shown in Figures ~\ref{fig:RF_pot} and \ref{fig:DC_pot}.  The insets in each figure show a close-up of the potentials near the trap center and the residuals $V_\textrm{RF}(s,z) - V_\textrm{CH}(s,z)$ and $U_\textrm{static}(s,z) - U_\textrm{CH}(s,z)$ for the RF and static potentials, respectively.  The trap center is located at $R\approx430$~$\mu$m; the normalization coefficients are $\ell_\textrm{RF}\approx413$~$\mu$m and $\ell_\textrm{static}\approx328$~$\mu$m.  The optimized parameters are listed in Table~\ref{table1}.

\begin{table}[htdp]
\caption{The optimized values for both the geometric parameters and the static potentials.}
\begin{center}
\begin{tabular}{|c|c||c|c|}
\hline
Parameter & Value & Parameter & Value\\
\hline
\hline
$A_h$ & 0.676 & $V_0$ & 1~V\\
\hline
$K_h$ & 1.68 & $U_0$ & -42.97~V\\ 
\hline
$V_h$ & 2.06 & $U_1$ & 1.09~V\\
\hline
$\theta_h$ & $16.7^\circ$ &$U_2$ & 1.03~V \\
\hline
\end{tabular}
\end{center}
\label{table1}
\end{table}

Both the RF and static potentials have large zones near the trap center that approximate very well (to greater than 95\%) the harmonic potential models.  Although it may be possible to further optimize the potentials by using custom-fabricated electrode structures, the potentials generated by stock electrode parts should be of sufficient quality for quantum information and Coulomb crystal experiments.

\section{Crystal Phase Transition}

\begin{figure}[htbp]
\centering
\includegraphics[width=8cm,keepaspectratio]{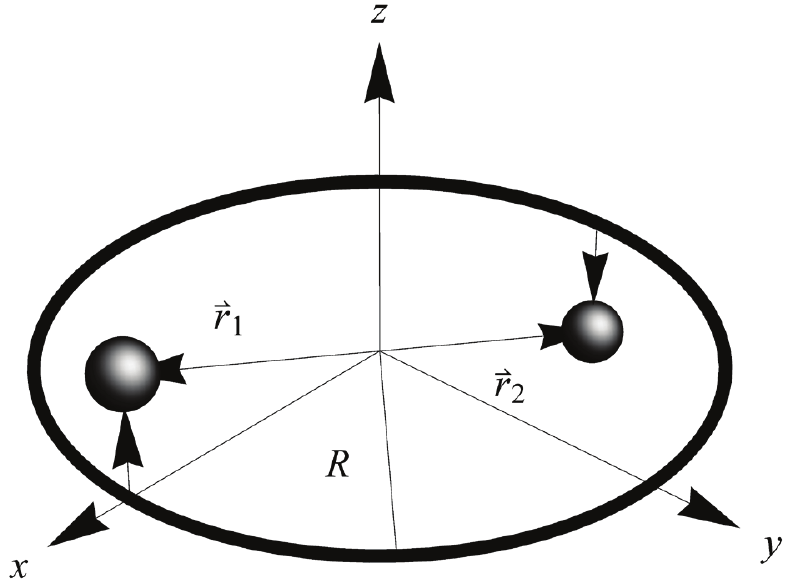}
\caption{Two ions in the halo trap begin on opposite sides of the ring.  There is a second-order phase transition where the ions shift off of the ring in opposite directions.}
\label{fig:phase}
\end{figure}

In this final section we derive the first non-trival ion crystal phase transition for two ions in the compact halo trap.  Whereas the first non-trivial phase transition for ion crystals in a linear trap occurs with three ions in an anisotropic trap \cite{rafac},  there is a second-order phase transition for two ions in the halo trap between the crystal configuration where both ions lie in the $z=0$ plane (Fig.~\ref{fig:phase}) and the regime where both ions are located at stable equilibrium positions equidistant from the $z=0$ plane as shown in the figure.  In deriving the condition for the phase transition, we follow the derivation of Schweigert, et al. \cite{schweigert} with the key difference that we consider a harmonic trap in both the $r$ and $z$ diminsions, allowing the ions to move in three-dimensional space.

We consider a three-dimensional system of identically charged particles of mass $m$ and charge $q$ confined to our halo ring by a ring-shaped external potential $V_r = (1/2) m \omega_r^2 (r-R)^2$ (with trap radius $R$ as above) and a second, independent, harmonic potential in the $z$-direction $V_z=(1/2) m \omega_z^2 z^2$.  The interaction Hamiltonian, including Coulomb repulsion, of this classical system of $N$ particles is thus
\begin{align}
\label{eqUtot} H =& \sum_{i < j=1}^{N} \frac{q^{2}}{4\pi \epsilon_{0}} \left( \frac{1}{\left| \vec{r}_{i} - \vec{r}_{j} \right|} \right) + \sum_{i = 1}^{N} \frac{1}{2} m \omega_{r}^{2}\left(\left| \vec{r}_{i} \times \hat{z} \right| - R \right)^{2} \nonumber \\ &+ \sum_{i = 1}^{N} \frac{1}{2} m \omega_{z}^{2}\left( \vec{r}_{i} \cdot \hat{z} \right)^{2},
\end{align}
for each particle located at $\vec{r}_i$.  Following the convention in \cite{schweigert}, we re-scale the length and energy units in order to work with dimension-less parameters.  The length scale is defined as the ratio of the Coulomb and radial trap coupling constants,
\begin{equation}
r_{\star} \equiv \left( \frac{q^{2}}{4 \pi \epsilon_{0}}\frac{2}{ m \omega_{r}^{2}} \right)^{1/3}.
\end{equation}
The length scales for a number of different common ion trap species are listed in Table~\ref{table2}.  All of these are similar in scale the model compact halo trap radius.  We also define an energy scale based on the radial potential and this length scale, $E_\star \equiv (1/2) m \omega_r^2 r_\star^2$ and scale the overall energy by this parameter.  Finally, we define the trap aspect ratio as the ratio of the $z$ and $r$ trap frequencies, similar to a linear trap,
\begin{equation}
\alpha \equiv \frac{\omega_z^2}{\omega_r^2}.
\end{equation}

\begin{table}[htdp]
\caption{Various scaled radii for common ion trap configurations.  The final entry is for 300 nm diameter polystyrene spheres, charged to $1.6\times10^{-16}$~C.}
\begin{center}
\begin{tabular}{|c|c|c|}
\hline
Ion & $\omega_r/2\pi$ & $r_\star$\\
\hline\hline
$^{24}$Mg$^+$ &  2 kHz & 419 $\mu$m \\
\hline
$^{40}$Ca$^+$&  1.5 kHz & 428 $\mu$m \\
\hline
$^{171}$Yb$^+$&  0.8 kHz & 401 $\mu$m \\
\hline
300 nm PS &  100 Hz & 429 $\mu$m \\
\hline
\end{tabular}
\end{center}
\label{table2}
\end{table}%

\begin{figure}[htbp]
\centering
\includegraphics[width=8cm,keepaspectratio]{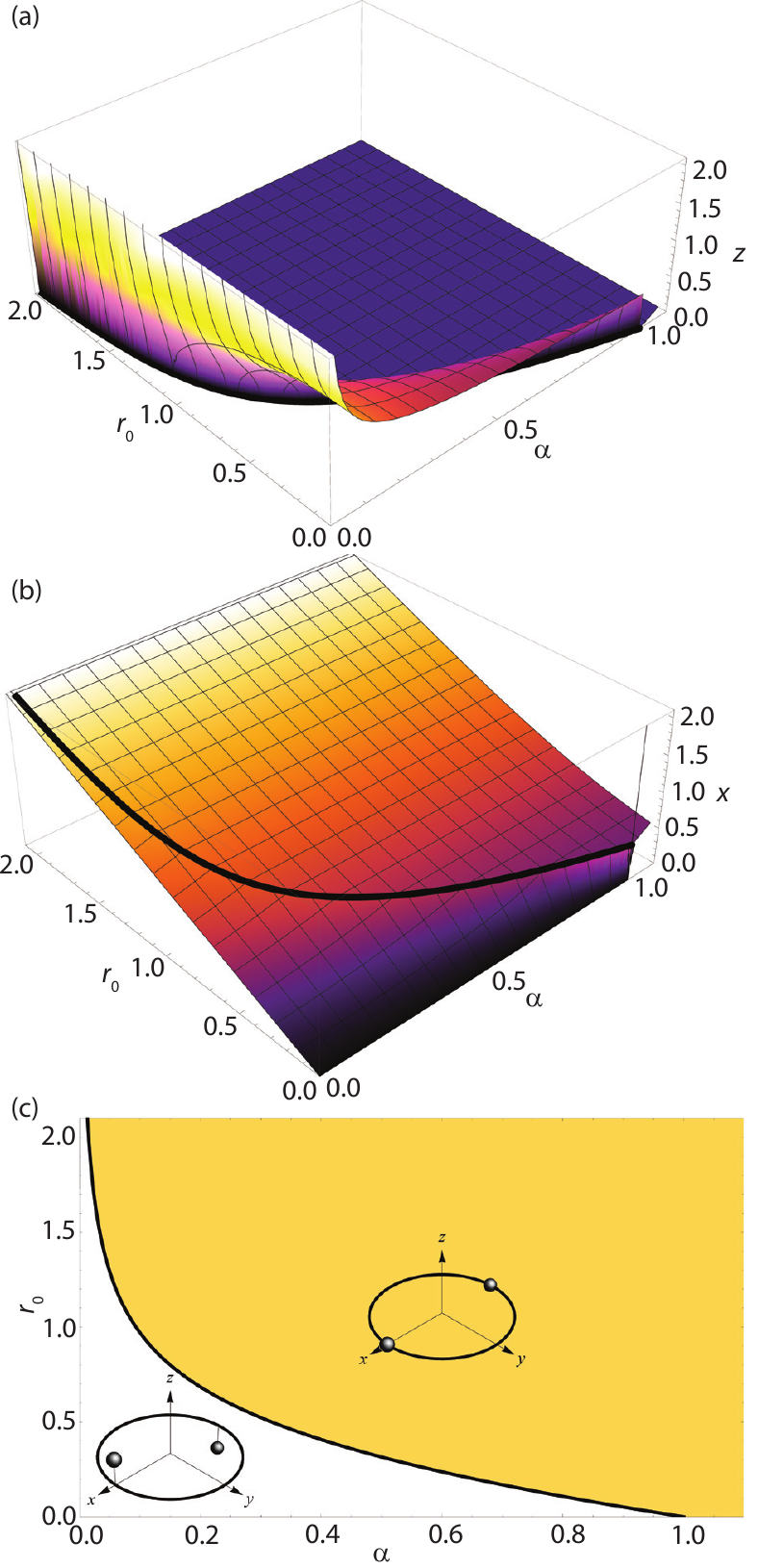} 
\caption{(Color Online) (a) The scaled $z$ position as a function of $\alpha$ and $r_0$.  The phase transition occurs along the line described by Eq.~(\ref{eqn:phase}). (b) The scaled $x$ position also shows the phase transition. (c) The phase transition in $\alpha$ and $r_0$ space.}
\label{fig:r_alpha}
\end{figure}

In the simplest configuration, the ions all lie in the $z=0$ plane and distribute themselves equally around a circle \cite{lupinski} with their center of mass located at the center of the ring $x_{cm}=y_{cm}=z_{cm}=0$.  Since no external forces act on the ions in this trap, the center-of-mass of the system will not move. We use this to simplify the ion locations for the $N=2$ case, making the assumption that the ions start on opposite ends of the $x$-axis, the positions of the two ions will be related by the following three constraints: 
\begin{align}
x_{1} &= -x_{2} \nonumber \\
y_{1} &= y_{2} = 0 \nonumber \\
z_{1} &= -z_{2}.
\end{align}
These assumptions reduce Eq.~(\ref{eqUtot}) to 
\begin{equation}
\label{eqUtotIon2_2} H =  \frac{1}{2\sqrt{ x^{2} + z^{2}}} + 2 \alpha z^{2} + 2\left( \left| x \right| - r_0 \right)^{2}.
\end{equation}
where we have scaled the distances, $x = x_1/r_\star$, $z = z_1/z_\star$, and energy $H\rightarrow H/E_\star$ as noted above.  The trap center is also scaled by the same parameter, $r_0 = R/r_\star$.  This unitless radius now describes the ratio of the physical trap $R$ to the strength of the Coloumb interaction.  We will assume that the trap is small enough that the ions interact strongly across the diameter of the trap circle.

The equilibrium positions of the two ions are found from the first spatial derivatives of the interaction Hamiltonian in the $x$ and $z$ directions,
\begin{align}
\frac{\partial H}{\partial x} = & 0 \nonumber \\
\frac{\partial H}{\partial z} = & 0.
\end{align}
The first set of stable equilibrium positions are at
\begin{align}
x=&\frac{r_0}{3}+\frac{16^{1/3} r_0^2}{3 \left(27+16 r_0^3-3 \sqrt{81+96 r_0^3}\right)^{1/3}} \nonumber \\ &+\frac{\left(27+16 r_0^3-3 \sqrt{81+96 r_0^3}\right)^{1/3}}{432^{1/3}} \nonumber \\
z=&0.
\end{align}
This solution is independent of the trap aspect ratio $\alpha$ and is valid for  
\begin{equation}
r_0>\frac{\left|\alpha - 1\right|}{2\alpha^{1/3}}\;\textrm{and}\; \alpha>1. 
\label{eqn:phase}
\end{equation}
As the ring becomes small, or the trap aspect ratio decreases, corresponding to a weakening of the $z$ potential, the ions reach a point where they shift off of the $z=0$ plane into the second crystal phase described by the equilibrium positions 
\begin{align}
x&=\frac{r_0}{\left|\alpha - 1\right|} \nonumber\\
z&=\frac{1}{2} \sqrt{\frac{(\alpha-1)^2-4 r_0^2 \alpha ^{2/3}}{(\alpha-1)^2 \alpha ^{2/3}}}.
\end{align}
The $x$ and $z$ positions as a function of $r_0$ and $\alpha$ are shown in Fig.~\ref{fig:r_alpha}(a-b).  This is a second-order phase transition since the derivative of the mean radial position of the ion cloud is discontinuous.  The boundary between the two phases is also shown in Fig.~\ref{fig:r_alpha}(c).

We have shown that there is a non-trivial phase transition for two ions in the compact halo trap.  Future work includes solving for the phase transition for larger numbers of ions, in particular for odd numbers of ions, which have a different behavior for small numbers of ions \cite{schweigert}.  We have also presented an optimized electrode configuration for creating the compact halo trap using readily-available electrode parts.  This type of trap is an alternative topology to more common point and linear ion traps and may be useful as a small, novel trap geometry for some quantum information applications.

\bibliographystyle{unsrtnat}

\end{document}